\documentclass{PoS}

\newcommand{\order}{{\it O}}
\newcommand{\ba}{\begin{eqnarray}}
\newcommand{\ea}{\end{eqnarray}}
\newcommand{\be} {\begin{equation}}
\newcommand{\ee} {\end{equation}}

\title{Kaon semileptonic form factors with $N_f=2+1+1$ HISQ fermions and physical light 
quark masses}

\ShortTitle{$K$ semileptonic form factor with HISQ fermions at the physical point}

\author{
\speaker{E.~G\'amiz}$^{a}$,
A.~Bazavov$^b$,\thanks{Present address: Department of Physics and Astronomy, University of 
Iowa, IA, USA}
C.~Bernard$^c$,
C.~Bouchard$^d$,
C.~DeTar$^e$,
D.~Du$^f$,
A.X.~El-Khadra$^f$,
J.~Foley$^e$,
E.D.~Freeland$^g$,
Steven~Gottlieb$^h$,
U.M.~Heller$^i$,
J.~Kim$^j$,
A.S.~Kronfeld$^k$,
J.~Laiho$^{l}$\thanks{Present address: Department of Physics, Syracuse University, 
Syracuse, NY, USA},
L.~Levkova$^e$,
P.B.~Mackenzie$^k$,
E.T.Neil$^k$\thanks{Present address: Department of Physics, University of Colorado, Boulder, CO, USA 
and RIKEN-BNL Research Center, Brookhaven National Laboratory, Upton, NY, USA},
M.B.~Oktay$^e$,
Si-Wei Qiu$^e$,
J.N.~Simone$^k$,
R.~Sugar$^m$,
D.~Toussaint$^j$,
R.S.~Van~de~Water$^k$,
and
Ran Zhou$^h$\thanks{Present address Fermi National Accelerator Laboratory, Batavia, IL, USA}
 \\ \\
\llap{$^a$}
CAFPE and Departamento de F\'{\i}sica Te\'orica y del Cosmos,
Universidad de Granada,\hspace*{-0.4em}\thanks{Supported in part by the MINECO, Junta 
de Andaluc\'{\i}a, and European Commission.}~Granada, Spain \\
\llap{$^b$}Physics Department, Brookhaven National Laboratory,\hspace*{-0.4em}
    \thanks{Operated by Brookhaven Science Associates, LLC, under
    Contract No.\ DE-AC02-98CH10886 with the US DOE.}~
Upton, NY, USA \\
\llap{$^c$}Department of Physics, Washington University, St.~Louis, MO, USA \\
\llap{$^d$}Department of Physics, The Ohio State University, Columbus, Ohio, USA\\
\llap{$^e$}Physics Department, University of Utah, Salt Lake City, UT, USA \\
\llap{$^f$}Physics Department, University of Illinois, Urbana, IL, USA \\
\llap{$^g$}Liberal Arts Department, School of the Art Institute of Chicago, Chicago, Illinois, USA\\
\llap{$^h$}Department of Physics, Indiana University, Bloomington, IN, USA \\
\llap{$^i$}American Physical Society, One Research Road, Ridge, NY, USA \\
\llap{$^j$}Department of Physics, University of Arizona, Tucson, AZ, USA \\
\llap{$^k$}Fermi National Accelerator Laboratory,\hspace*{-0.4em}
    \thanks{Operated by Fermi Research Alliance, LLC, under Contract
    No.~DE-AC02-07CH11359 with the US DOE.}~
Batavia, IL, USA \\
\llap{$^l$}SUPA, School of Physics \& Astronomy, University of Glasgow,
Glasgow, UK\\
\llap{$^m$}Department of Physics, University of California, Santa Barbara,
USA\\

E-mail: \email{megamiz@ugr.es}}
\author{Fermilab Lattice and MILC Collaborations\\
}

\abstract{
We present results for the form factor $f_+^{K\pi}(0)$, needed to extract the CKM matrix element 
$|V_{us}|$ from experimental data on semileptonic $K$ decays, on the HISQ $N_f=2+1+1$ MILC 
configurations. The HISQ action is also used for the valence sector. The data 
set used for our final result includes three different values of the lattice spacing and data at the 
physical light quark masses. We discuss the error budget and how this calculation improves 
on our previous determination of $f_+^{K \pi}(0)$ on the asqtad $N_f=2+1$ MILC configurations. 
}

\FullConference{31st International Symposium on Lattice Field Theory - LATTICE 2013\\
		July 29 - August 3, 2013\\
		Mainz, Germany}

\begin{document}

\section{Motivation}

A precise determination of the CKM parameter $\vert V_{us}\vert$ has been the subject  
of extensive work using $K$ leptonic and semileptonic decays, 
as well as hadronic $\tau$ decays. 
The goal is to test the unitarity of the CKM matrix in the first row and establish 
stringent constraints on the scale of the new physics that could contribute to 
these processes~\cite{Antonellietal2010,Cirigliano10}. 

In Ref.~\cite{Ktopilnu_HISQ} we present our result for the $K$ semileptonic form factor 
$f_+(0)$, which includes for the first time data at the physical light quark masses. 
In this contribution we present further details on the chiral interpolation and continuum 
extrapolation as well as on our study of the other systematic errors that enter our result. 
Our result for $f_+(0)$ can be used together with experimental data on exclusive 
semileptonic $K$ decays to extract $|V_{us}|$ with a precision that is currently 
limited by the uncertainty in $f_+(0)$ \cite{Bazavov:2012cd,Ktopi}.  
The form factor is
\ba
\langle \pi \vert V^\mu \vert K\rangle =  f_+(q^2) \left[p_K^\mu 
+ p_\pi^\mu - \frac{m_K^2-m_\pi^2}{q^2}q^\mu\right]+ f_0(q^2)\frac{m_K^2-m_\pi^2}{q^2}q^\mu\,.
\ea
The set-up of our calculations is described in Refs.~\cite{Bazavov:2012cd} 
and~\cite{Lattice2012}. We obtain $f_+(0)$ from the relation 
$f_+(0)=f_0(0) = \frac{m_s-m_l}{m_K^2-m_\pi^2}\langle \pi(p_\pi)\vert S \vert K(p_K)\rangle$ 
and simulate directly at zero momentum transfer, $q^2\approx0$, by tuning the external momentum 
of the $\pi$ using partially twisted boundary conditions. Unlike in our asqtad $N_f=2+1$ 
calculation~\cite{Bazavov:2012cd}, here we do not include correlation functions with moving 
$K$'s since they are considerably noisier than with moving $\pi$'s~\cite{Lattice2012}. 

We use the HISQ action for the sea and valence quarks, simulating 
on the HISQ $N_f=2+1+1$ MILC configurations~\cite{HISQensembles}. We analyze 
the ensembles listed in Table~\ref{tab:ensemblesHisq}, although we use the ensemble 
with the smallest lattice spacing, $a\approx 0.06~ {\rm fm}$, only as a consistency check. 
The $a\approx 0.12~{\rm fm}$ ensemble with $m_\pi L=5.36$ is used only for an estimate of  
finite-volume (FV) effects. In order to avoid autocorrelations, we block our data by four. We 
try different blocking sizes and find that the results from the correlator fits, both 
central values and errors, stabilize when the data is blocked by four. We already discussed the 
correlator fit strategy and the fit functions used in Refs.~\cite{Bazavov:2012cd} 
and~\cite{Lattice2012}, so we do not repeat that here. We just show the results for $f_+(0)$ 
for the ensembles we include in our main analysis in Table~\ref{tab:f+ensembles} and 
in Fig.~\ref{fig:ChPTcentral}. Statistical (bootstrap) errors are $\sim0.2$--0.4\%. 
We observe that the variation of the results with 
lattice spacing is less than the statistical errors, except for the ensemble 
with $a\approx 0.15~{\rm fm}$.

\begin{table}[th]

\vspace*{-0.3cm}
\begin{center}\begin{tabular}{c|ccc|ccccccc}
\hline\hline
      $\approx a$(fm) & $am_l^{\rm sea}$ & $am_s^{\rm sea}$ & $am_c^{\rm sea}$ & 
$am_s^{\rm val}$ & $m_{\pi}^P$ &  $m_{\pi}^{{\rm RMS}}$ & $L({\rm fm})$ & $m_\pi L$ 
& $N_{\rm conf}$ & $N_{{\rm s}}$ \\
\hline
0.15  & 0.00235 & 0.0647 & 0.831 & 0.06905 & 133 & 311 & 3.2 & 3.30 & $1000$ & $4$   \\
\hline
0.12 & 0.0102 & 0.0509 &  0.635 & 0.0535 & 309 & 370 & 3.00 & 4.54 &  $1053$ & $8$ \\
     & 0.00507 & 0.0507 & 0.628 & 0.053 & 215 & 294 & 3.93 & 4.29 &   $993$ & $4$ \\
     & 0.00507 & 0.0507 & 0.628 & 0.053 & 215 & 294 & 4.95 & 5.36 &   $391$ & $4$  \\
     & 0.00184 & 0.0507 & 0.628 & 0.0531 & 133 & 241 & 5.82 & 3.88 &  $945$ & $8$  \\
\hline
0.09   & 0.0074 & 0.037 & 0.440 & 0.038 & 312 & 332 & 2.95 & 4.50 &   $775$ & $4$ \\
       & 0.00363 & 0.0363 & 0.430 & 0.038 & 215 & 244 & 4.33 & 4.71 &  $853$ & $4$ \\
       & 0.0012 & 0.0363 & 0.432 &  0.0363 & 128 & 173 & 5.62 & 3.66 &  $621$ & $4$ \\
\hline
0.06   & 0.0048 & 0.024 & 0.286 & 0.024 & 319 & 323 & 2.94 & 4.51 & $362$ & $4$ \\
\hline\hline
\end{tabular}\end{center}
\vspace*{-0.3cm}
\caption{Parameters of the $N_f=2+1+1$ gauge-field ensembles used in
this work and details of the correlation functions generated.
$N_{\rm conf}$ is the number of configurations included in our analysis,
$N_{{\rm s}}$ the number of time sources used on each configuration,
and $L$ the spatial size of the lattice. $m_\pi$'s are given
in MeV, where $m^P_\pi$ is the Goldstone (pseudoscalar taste) $\pi$ mass and 
$m^{{\rm RMS}}_\pi$ the root-mean-squared (over all tastes) $\pi$ mass. 
\label{tab:ensemblesHisq}}
\end{table}

\begin{table}[tb]
\begin{center}
\begin{tabular}{cccc}
\hline\hline
$\approx a ({\rm fm})$ & $m_\pi^P\approx m_\pi^{{\rm phys}}$ & $m_l=0.1m_s$ & $m_l=0.2m_s$ \\
\hline
0.15 & 0.9744(24) & - & - \\
0.12 & 0.9707(18) & 0.9808(22) & 0.9874(24) \\
0.09 & 0.9699(36) & 0.9807(22) & 0.9868(18) \\
\hline\hline
\end{tabular}
\caption{Values of $f_+(0)$ included in the chiral interpolation and continuum 
extrapolation. Errors are statistical only, from 500 bootstrap ensembles. 
\label{tab:f+ensembles} }
\end{center}

\vspace*{-0.7cm}
\end{table}

\section{Chiral-continuum interpolation/extrapolation} 

Although we have data at the physical (and smaller) light-quark masses, 
we also include data from ensembles with larger light-quark masses in our analysis 
(see Table~\ref{tab:ensemblesHisq}), and hence use chiral perturbation theory 
($\chi$PT) to interpolate to the physical point. This allows us to correct for small 
mistunings of quark masses and reduce statistical errors. Due to the Ademollo-Gatto (AG)  
theorem, $f_+(0)$ is constrained to follow the chiral 
expansion $f_+(0)=1+f_2+f_4+f_6 + \dots$, with $f_{2i}$ chiral corrections of $\order(p^{2i})$ 
that go to zero in the $SU(3)$ limit as $(m_K^2-m_\pi^2)^2$ up to discretization errors of 
$\order(\alpha_S²a^2,a^4)$~\cite{Ktopilnu_HISQ}. Following the same strategy as in our 
asqtad $N_f=2+1$ analysis, we use partially quenched staggered 
$\chi$PT (PQS$\chi$PT) at NLO~\cite{KtopilnuSChPT} plus regular continuum $\chi$PT at 
NNLO~\cite{BT03}, and add the discretization effects mentioned above and  the dominant 
$a^2$ corrections that respect the AG theorem (other than those included explicitly 
by the NLO PQS$\chi$PT), {\it i.~e.}, $\order\left((m_K^2-m_\pi^2)^2\alpha_s a^2, 
(m_K^2-m_\pi^2)^2\alpha_s^2 a^2\right)$ terms. We also include isospin  
corrections at NLO from Ref.~\cite{Gasser:1984ux} and interpolate to the physical $\pi$ 
and $K$ masses but with electromagnetic effects removed~\cite{FLAG2,Basak:2013iw},  
{\it i.~e.}, $m_{\pi^+}^{QCD}=135.0~{\rm MeV}$,
$m_{K^0}^{QCD}\approx m_{K^0}^{{\rm phys}}=497.7~{\rm MeV}$, and 
$m_{K^+}^{QCD}=491.6{\rm MeV}$. $m_{K^+}^{\rm QCD}$ above is used only in $f_2$, 
to account for the leading isospin corrections. The fit function is

\vspace*{-0.5cm}
\ba\label{eq:ChPTtwoloop}
f_+(0)  =   1   & + &   f_2^{{\rm PQS\chi PT}}(a) + K_1\,\sqrt{r_1^2a^2
\bar\Delta\left(\frac{a}{r_1}
\right) ^2} + K_3\,\left(\frac{a}{r_1}\right)^4 + f_4^{{\rm cont.}}\nonumber\\ 
 & + &   r_1^4(m_\pi^2-m_K^2)^2 \left[\,C_6+K_2\,\sqrt{r_1^2a^2\bar\Delta\left(\frac{a}{r_1}\right) ^2}
+ K_2'\,r_1²a^2\bar\Delta\right]\,,
\ea
where the constants $K_i$ and $C_i$ are fit parameters to be determined by the chiral 
fits using Bayesian techniques~\footnote{Notice that the fit parameters used here differ from 
the ones in Ref.~\cite{Ktopilnu_HISQ} by factors of $r_1$, although we are using the same 
notation for both sets.}, $\bar\Delta$ is the average taste splitting
$\bar\Delta = \frac{1}{16}\left(\Delta_P+4\Delta_A+6\Delta_T+4\Delta_V+\Delta_I\right)$, and 
$r_1^2 a^2 \bar \Delta$ is a proxy for $\alpha_s^2 a^2$. 
$C_6$ is proportional to the combination of low-energy constants (LEC's) 
$C_{12}+C_{34}-L_5^2$~\cite{p6Lagrangian}, where $C_{ij}$ are $\order(p^6)$ and $L_5$ is $\order(p^4)$. 
We follow the same approach as in Ref.~\cite{Bazavov:2012cd} and take the  $\order(p^4)$ LEC's 
and the taste-violating hairpin parameters~\cite{hairpins} as constrained fit parameters. 
The corresponding uncertainties are thus included in the error of the central value. 
The HISQ taste splittings, which we take from Ref.~\cite{HISQensembles}, are known 
precisely enough that their error has no impact on the final uncertainty. 
The prior central values and widths 
we use in our fits are in Table~\ref{tab:priors}. 

\begin{table}[tb]
\vspace*{-0.5cm}
    \centering
\caption{
Priors (central value$\pm$width) for the fit parameters entering in Eq.~(2.1).   
The $\chi$PT parameter $s$ is given by $1/(16\pi^2(r_1f_\pi)^2)$. 
The priors listed for the hairpin parameters are for the $a\approx 0.12~{\rm fm}$ 
ensembles, and those for the other lattice spacings are obtained by rescaling this number,
assuming that the hairpin parameters scale like the $\Delta_\Xi$. 
The central values for the NLO LEC's $L_i's$ are from fit 10 in Ref.~\cite{LisHans}. 
 \label{tab:priors}}
\begin{center}  
\begin{tabular}{|cc|ccccc|}
\hline
\hline
  $r_1^2a^2\delta_V^{{\rm HISQ}}$ & $r_1^2a^2\delta_A^{{\rm HISQ}}$ &  $K_1$ & $K_2$ & $K_2'$ & $K_3$ &
$C_{2i}$  \\
\hline
$0.057\pm0.033$&$-0.0782\pm0.0040$& $0\pm 0.01$ & $0\pm 0.03$ 
&  $0\pm 0.081$ &  $0\pm 0.015$ &  $0\pm s^{i-1}$ \\
\hline\hline
\end{tabular}
\begin{tabular}{|ccccc|}
  $L_1^r(M_\rho)$ &  $L_2^r(M_\rho)$ &  $L_3^r(M_\rho)$ &  $L_4^r(M_\rho)$ &  $L_5^r(M_\rho)$ \\
\hline
$0.43\pm0.24$&$0.73\pm0.24$&$-2.30\pm0.74$&$0.0\pm0.6$&$0.97\pm0.22$\\
\hline\hline
\end{tabular}
\begin{tabular}{|ccc|}
$(2L_6^r-L_4^r)(M_\rho)$&  $L_7^r(M_\rho)$ &  $L_8^r(M_\rho)$ \\
\hline
$0.0\pm0.4$&$-0.31\pm0.28$&$0.60\pm0.36$\\   
\hline\hline
\end{tabular}
\end{center}

\vspace*{-1.cm}
\end{table}

With the fit function in Eq.~(\ref{eq:ChPTtwoloop}) and including the data at 
$a\approx 0.15,0.12,0.09~{\rm fm}$ in Table~\ref{tab:ensemblesHisq} we get
$f_+(0)=0.9703(23)$. The interpolation as well as the data points included in the fit and those 
used for estimating systematic errors or as a consistency check, 
are shown in Fig.~\ref{fig:ChPTcentral}. 

\subsection{Systematic error analysis}

\label{sec:errors}

As explained in Ref.~\cite{Ktopilnu_HISQ}, we expect that the error in 
our chiral-continuum fit value includes both statistical and 
discretization errors. 
In order to check this expectation, we also follow an
alternate strategy to try to separate statistical from discretization errors.
The central value for this second strategy is given by a fit that does not include
any extra $a^2$ terms (besides those in the one-loop S$\chi$PT expression),
but without including the $a\approx 0.15~{\rm fm}$ point, $f_+(0)=0.9708(15)$.
Then we perform a number of fits using fit functions in which we
parametrize discretization errors in different ways, including all possible
combinations of the four terms in Eq.~(\ref{eq:ChPTtwoloop}) and continuum NNLO 
$\chi$PT plus analytic $a^2$, $\alpha_s a^2$, and $a^4$ terms.
The different parametrizations do not move the central value more than 0.0010, well
below the statistical error. If we take this variation as the estimate of discretization 
errors for this alternate fit strategy, we obtain a combined statistical and 
discretization error of $\pm 0.0018$, which is smaller than the corresponding error 
for our central result.

The chiral interpolation is very much constrained by the data at the 
physical light quark masses, so the dependence of the fit result 
on the $\chi$PT parameters is very much suppressed. For example, we can test
the choice of fit function by using an analytic parametrization instead of the
continuum two-loop ChPT expression. The results from NNLO, N${}^3$LO, and N${}^4$LO 
analytic parametrizations agree with our central value well within statistical
errors.

A more accurate test of higher order effects in the chiral expansion is achieved by
adding N${}^3$LO, and N${}^4$LO analytic terms to the fit function. Adding an N${}^3$LO  
term $C_8\,(m_K^2-m_\pi^2)^2m_\pi^2$ to Eq.~(\ref{eq:ChPTtwoloop}) with an unknown 
but constrained coefficient slightly changes the central value to 0.9704 and 
increases the error to 0.0024~\footnote{This increase in the error is 
a measure of the chiral interpolation error. Another ways of estimating this error, 
such as replacing $f_\pi$ by $f_K$ or the chiral limit of the decay constant, $f_0$, 
at NNLO, give similar results.} When, in addition, we add a $N^4LO$ term,
$C_{10}\,(m_K^2-m_\pi^2)^2m_\pi^4$, the central value and error do not change. In other words,
the result from the chiral and continuum fit stabilizes once we include up to $N^3LO$
chiral corrections. We thus take the result from that fit, $f_+(0)=0.9704(24)$,
as our central value for the form factor and the error including statistics,
discretization effects, and higher order chiral corrections.

\begin{center}
\begin{figure}[ŧh]
\begin{center}
\vspace*{-1.3cm}
\includegraphics[width=0.6\textwidth]{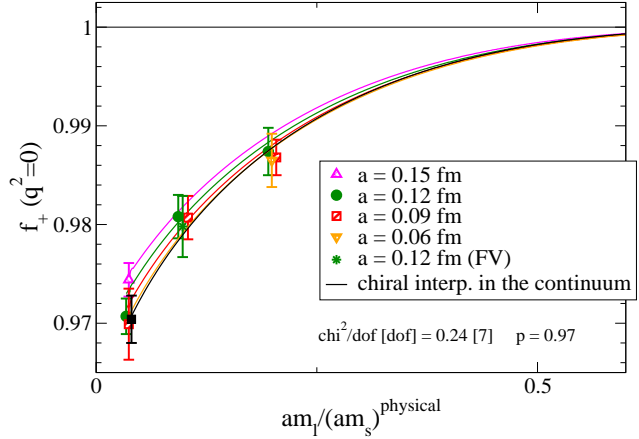}
\end{center}
\vspace*{-0.5cm}
\caption{Form factor $f_+(0)$ {\it vs}.~light-quark mass. Errors shown on the data
points are statistical only, obtained from 500 bootstrap ensembles.
Different symbols and colors denote different lattice spacings, and the corresponding
colored lines show the chiral interpolation at fixed lattice spacing.
The green star labels the ensemble we use to estimate FV effects.
The solid black line is the interpolation in the light-quark mass, keeping $m_s$
equal to its physical value, and turning off all discretization effects.
\label{fig:ChPTcentral} }

\vspace*{-0.5cm}
\end{figure}
\end{center}

We have, however, another systematic effect arising from the fact that for some 
ensembles we have different strange-quark masses in the sea and in the valence sectors.
This difference is treated correctly at NLO, since we have a partially quenched S$\chi$PT fit
function, but at NNLO the continuum expression is only evaluated for the full QCD
case, with no difference between the sea and valence sectors. We use
$m_s^{\rm val}$ in the NNLO piece, $f_4$, for our central result. But, in order to
estimate the uncertainty associated with this choice, we redo the fit replacing
$m_s^{{\rm val}}$ by  $m_s^{{\rm sea}}$ in $f_4$ except in the overall factor $(m_K²-m_\pi^2)^2$
which is generated by the valence sector. The shift in $f_+(0)$ is 0.0003 for
our main analysis, which we take as the associated systematic error.

Finite volume effects can be systematically addressed in the framework of $\chi$PT,
replacing the infinite volume integrals by FV ones and extrapolating to 
the infinite volume limit. The S$\chi$PT incorporating these effects for 
our calculation of $f_+(0)$ with partially twisted boundary conditions is  
not yet available, although work is in progress~\cite{Ktopi_FV}.
In order to estimate the FV error, we perform two tests. We carried out an
additional simulation on an ensemble with the same parameters as the $a \approx   
0.12~ {\rm fm}$, $m_l=0.1m_s$ but with a larger volume (fourth line in
Table~\ref{tab:ensemblesHisq} and open circle in Fig.~\ref{fig:ChPTcentral}). 
This larger volume simulation gives a result 
$\sim 0.1\%$ lower, or about half of the smaller of the two statistical errors of 
the ensembles we are comparing. We check the stability of this shift by 
performing a variety of correlator fits with different parameters without finding a 
larger effect. We also perform a second
test in which we replace the logarithmic functions and their derivatives in the NLO 
chiral expression by their FV counterparts~\cite{Bernard:2001yj}, 
$\textrm{ln}\frac{m^2}{\Lambda^2}\rightarrow\left(\textrm{ln}\frac{m^2}{\Lambda^2}
+\delta_1(mL)\right)$ and  $-\left(\textrm{ln}\frac{m^2}{\Lambda^2} + 1\right)\rightarrow
 -\left(\textrm{ln}\frac{m^2}{\Lambda^2} + 1\right) + \delta_3(mL)$, 
and redo the chiral interpolation and continuum (+infinite volume) extrapolation. 
With this replacement $f_+(0)$ decreases by $0.11\%$.
This test does not take into account all the possible FV corrections or
the fact that we are using twisted boundary conditions (which modifies the FV
integrals), but it gives us an idea of the size of these corrections. We take  
the full size of the statistical error of the $a\approx0.12~{\rm fm}$, $m_l=0.1m_s$ 
ensemble, $0.2\%$, as our FV error estimate. We consider the effect of the scale uncertainty 
on the dimensionless quantity $f_+(0)$. Here we use $r_1=0.3117\pm0.0022~{\rm fm}$ from 
Ref.~\cite{decayconstants2011}, which yields an error of $\pm0.0008$ on $f_+(0)$. 
Finally, for the estimate of the higher order isospin corrections in the 
$K^0\pi^+$ mode, we take twice the difference between the isospin-conserving 
and isospin-violating calculation of $f_+(0)$ at NNLO from Ref.~\cite{Bijnens:2007xa}.

\section{Conclusions}

Our final result for the vector form factor is
\ba
f_+(0)=0.9704(24)(22)=0.9704(32)\,,
\ea
where the first error in the middle is the combined statistical, discretization, 
and chiral interpolation error, and the second is the sum in quadrature of the other 
systematic errors discussed above. Combining the two in quadrature again yields the 
error on the right. We discuss the implications of this result for the unitarity of 
the CKM matrix in Ref.~\cite{Ktopilnu_HISQ}.

The alternate fit strategy in which we try to disentangle statistical and discretization
errors, estimating the other systematic errors in the same way we do for our
main strategy gives the result: $f_+(0)=0.9708(15)(24)=0.9708(28)$, where the first
error is statistical plus higher order terms in the chiral expansion, and the
second the remainder of the systematic errors, including discretization effects. The
total error of this second strategy is  slightly smaller than the one in our
main analysis, which confirms the robustness of our systematic error analysis.

We also perform a combined analysis of our HISQ $N_f=2+1+1$ data and the asqtad 
$N_f=2+1$ data analyzed in Ref.~\cite{Bazavov:2012cd}. We use the fit function in 
Eq.~(\ref{eq:ChPTtwoloop}) plus N${}^3$LO and N${}^4$LO chiral corrections and the 
one in Eq.~(4.2) of Ref.~\cite{Bazavov:2012cd} (again, plus N${}^3$LO and N${}^4$LO 
chiral corrections) for the HISQ on HISQ and HISQ on asqtad data, 
respectively.\footnote{Although in Ref.~\cite{Bazavov:2012cd} we did not include the 
N${}^3$LO and N${}^4$LO chiral corrections in the fit function, the numerical difference 
of the fit results with and without those corrections is negligible within current 
precision.}  Notice that $f_2^{PQS\chi PT}(a)$ is different for the 
two sets of data, since the current analysis uses the HISQ action for both the sea and 
the valence quarks while the asqtad one is a mixed-action calculation 
with asqtad quarks in the sea and HISQ in the valence sector. The $PQS\chi PT$ expressions, 
which can be found for both cases in Ref.~\cite{KtopilnuSChPT}, take into account the 
differences between valence and sea as well as the particularities of the specific staggered 
action used. Among other parameters, the continuum LEC's and the coefficients $C_{2i}$ are 
the same for all data. Our combined fit provides an average of the two 
results taking into account correlations in a proper way. The result is 
$f_+(0) =  0.9686(17)(14)(6)(20)(2)=0.9686(30)\,$, 
where the first error is, again, the statistical+discretization+higher order chiral 
corrections error, the second one is from the mistuning of $m_s$ in the sea, the 
third reflects the uncertainty in $r_1$, the fourth is our estimate of FV 
corrections, and the last higher order isospin effects. The errors are estimated 
in the same way as described in Sec.~\ref{sec:errors} above.

The result presented here and in Ref.~\cite{Ktopilnu_HISQ} already constitutes 
the most precise determination 
of the vector form factor $f_+(0)$ and the first one to include simulations directly 
at the physical light-quark masses. However, to match the experimental uncertainty, 
we need to reduce the uncertainty on $f_+(0)$ further. Work is therefore continuing 
to address the two main sources of uncertainty in our result, statistics and FV effects. 
On one hand, there is an ongoing calculation of FV corrections at one-loop in 
S$\chi$PT~\cite{Ktopi_FV} which will allow us to eliminate part of this uncertainty 
and do a more reliable estimate of the remaining effect. On
the other hand, there are already more configurations in the ensembles that we have
analyzed and new ensembles that we plan to include in future work. Especially important
will be to reduce the statistical errors in the physical quark mass
$a\approx 0.09~{\rm fm}$ ensemble and to add an even finer lattice spacing at 
$a \approx 0.06~{\rm fm}$, also with physical masses.

\end{document}